\documentstyle[prl,aps,twocolumn]{revtex}
\begin{document}
\draft
 \title{Spin polarized tunneling in
ferromagnet/unconventional superconductor junctions}
\twocolumn[
\hsize\textwidth\columnwidth\hsize\csname@twocolumnfalse\endcsname
\author{Igor \v{Z}uti\'c and Oriol T. Valls}
\address{Department of Physics and Minnesota Supercomputer 
Institute,
University of Minnesota,
Minneapolis, Minnesota 55455}
\date{November 16, 1998}
\maketitle
\begin{abstract}
We study tunneling in ferromagnet/unconventional 
superconductor (F/S) junctions. We include the effects of spin 
polarization,  
interfacial resistance, and Fermi wavevector mismatch (FWM)
between the 
F and S regions. Andreev reflection (AR) at the F/S interface, governing 
tunneling
at low bias voltage, is strongly modified by these parameters. The 
conductance exhibits a wide variety of novel
features as a function of applied voltage.
\end{abstract}
\pacs{74.80.Fp, 74.50+r, 74.72-h}
]
Spin polarized transport and tunneling between ferromagnetic and 
superconducting materials has become a vigorously pursued area of research.
The studies performed\cite{soul,up,vas,dong,vas2} 
have implications
for the  understanding of unconventional superconductivty and
for the development  of devices using 
 spin polarized current\cite{soul,up,prinz}.
A recent study of a
ferromagnet/unconventional superconductor
(F/S) system has revealed\cite{vas2} a differential
conductance dip at zero bias (ZBCD), attributed to suppression
of Andreev reflection (AR) as  a
consequence of high spin polarization in the ferromagnet.
While there is substantial work\cite{soul,up,mers,been}
 on the interplay of ferromagnetism and
superconductivty in tunneling properties involving $s$-wave
 superconductors, there is still no adequate theory for these
 phenomena in the context of unconventional 
superconductivity.  Replacing an $s$-wave superconductor
by an unconventional one in a normal metal/superconductor (N/S) structure 
can drastically alter the conductance spectrum\cite{hu,xu,tan},
 and important
changes should also occur when such replacement is made in a F/S structure.
For N/S junctions,
zero bias conductance peaks (ZBCP) observed in HTSC's are
interpreted  as arising from
the sign change of the pair
potential  (PP) which leads to the
formation\cite{hu} of midgap surface states. The spectral
weight of these
states has, for a $d_{x^2-y^2}$ state,
a maximum for (110) oriented surfaces and vanishes
for (100)
surfaces. The absence of ZBCP in ${\rm Nd_{1.85} Ce_{0.15} Cu O_{4}}$  
is considered evidence\cite{xu} for a dominant  $s$-wave
component.

Previously published  work on tunneling in unconventional superconductors
examined the unpolarized case\cite{hu,tan}, or\cite{soul,up,been} F/S
junctions with an $s$-wave
PP\cite{ting}.
Here we consider the theory of tunneling spectroscopy 
for an F/S junction with arbitrary spin polarization.
We investigate the interplay of ferromagnetism and unconventional
superconductivity in forming  ZBCP, ZBCD,
and other features at finite bias.
We include the effects of the
exchange energy (related to the degree of 
polarization) interfacial barrier height and 
Fermi wavevector mismatch (FWM) in the
magnitude of the Fermi wavevectors in the F and S regions.
Variation of these parameters leads to rich behavior
and novel features in the conductance,
which require careful interpretation. 
Thus, FWM and spin
polarization can combine to yield a
ZBCP in an $s$-wave superconductor, while if one neglects FWM
\cite{soul,up,been,ting} the effect  of spin polarization 
invariably leads to suppression of AR.

We solve the Bogoliubov-de Gennes (BdG) equations\cite{been,hu,tan,bru}
for a ballistic F/S junction.
We extend the usual one-body Hamiltonian approach
of Ref. \onlinecite{been} to include: 1) scattering at the F/S interface, 
$x=0$, modeled by a potential $V({\bf r})=H
\delta(x)$, where
$H$ is the strength of the potential barrier, and 2) allow for FWM, i.e.
$E_F=\hbar^2 k_F^2/2m$ in the F region  at $x<0$ ( $E_F$ is the spin
averaged value,
$E_F=(\hbar^2 k_{F\uparrow}^2/2m+\hbar^2 k^2_{F\downarrow}/2m )/2$,) and
$E'_F=\hbar^2 k'^2_F/2m$ in the S region at $x>0$. 
We include the exchange energy\cite{been} $h({\bf r})=h_0 \Theta(-x)$, 
($\Theta(x)$ is a step function) and the pair
potential\cite{hu,tan} $\Delta({\bf k}', {\bf r})=\Delta({\bf
k}',{\bf r}) \Theta(x)$. From the invariance of the Hamiltonian with
respect to translations
parallel to $x=0$, the parallel component of 
the wave vector is conserved at the junction\cite{tan,bru}.
Then, the parallel component of the solutions of the BdG equations  
is a plane wave, and the problem reduces to a
1D one.

For an electron injected from the F side, with  spin $S=\uparrow,
\downarrow$, excitation energy $\epsilon$, and wavevector
${\bf k}^+_S$ at an angle $\theta$ from the interface
normal, there are (without spin-flip scattering)
four scattering processes \cite{tan,btk} with
different amplitudes.
For specular reflection at the interface, these are: 1)  Andreev 
reflection\cite{been,bru,nat}  with amplitude $a_S$ as a hole with 
spin, wavevector, and angle with the interface normal,
$\overline{S}$ (opposite to $S$),
${\bf k}_{\overline{S}}^-$, and $\theta_{\overline{S}}$,
respectively.
2) Ordinary
reflection with amplitude $b_S$ as
an electron with variables $S$, $-{\bf k}^+_{S}$, $-\theta$. 3)
Transmission with amplitude $c_S$ 
as an electronlike quasiparticle (ELQ) with ${\bf k}'^+_S$ 
and $\theta'_{S}$.
4) Transmission (amplitude $d_S$) as a holelike
quasiparticle  (HLQ) defined by $-{\bf k}'^-_S$, 
and $-\theta'_{S}$. Here we have used\cite{been,tan}
$k^{\pm}_{S} \approx k_{FS}$ and $k'^{\pm}_{S} \approx k'_{F}$, as explained
below.
The ELQ and HLQ have different, spin dependent, wavevectors, and
therefore they
feel different pair potentials $\Delta_{S+}$ and $\Delta_{S-}$, with
$\Delta_{S\pm}=|\Delta_{S \pm }| exp(i\phi_{S \pm})$.
The spin dependence (even without FWM)
of wavevectors and PP's is a novel feature of F/S junctions.

Conservation of $k_{\|S}$ yields the analogue of Snell's law,
$k_{FS}\sin\theta=k'_F\sin\theta'_S,  \quad S=\uparrow, \downarrow$,
and for  $k_{FS} >k'_F$ there is a spin dependent angle 
of total reflection.  
For a PP of the $d_{x^2-y^2}$ form,  and 
allowing for different angles $\alpha \in (-\pi/2, \pi/2)$ between the 
crystallographic $a$-axis and the interface normal, we have
$\Delta_{S\pm}=\Delta_0\cos(2 \theta'_{S\pm})$, where $\theta'_{S\pm}$
are related to 
$\theta'_S$ (and thus to the incident 
angle $\theta$, from ``Snell's law'') by 
$\theta'_{S\pm}=\theta'_{S}\mp\alpha$. The
spin dependence of these angles will produce more complicated 
conductance features.

In solving the BdG equations in the 
direction normal to the interface, we
have for the magnitude of the relevant
wavevectors: in the F region, $k^{\pm}_{S}=(2m/\hbar^2)^{1/2}
[E_F \pm \epsilon+\rho_S h_0]^{1/2}$,
where $\rho_S=\pm 1$ 
for $S=\uparrow$ $(\downarrow)$, and in the S region,
$k'^{\pm}_{S}=(2m/\hbar^2)^{1/2}
[E'_F\pm(\epsilon^2-|\Delta_{S\pm}|^2)^{1/2}]^{1/2}$.
In the regime 
of interest,  $E_F, E'_F $ $\gg$ $\max(\epsilon, |\Delta_{S\pm}|)$ and
\cite{been,tan,andre} we have $k^{\pm}_{S} \approx k_{FS}
\equiv(2m/\hbar^2)^{1/2}
[E_F +\rho_S h_0]^{1/2}$,
$k'^{\pm}_{S} \approx k'_F$.  From this and ``Snell's law'', the components
of  ${\bf k}^{\pm}_{S}$ and ${\bf k}'^{\pm}_{S}$ normal
and parallel to the interface can be found. We
write ${\bf k}^{\pm}_{S}\equiv(k_{S}, k_{\|S})$, and
${\bf k}'^{\pm}_{S}\equiv(k'_S, k_{\|S})$, 
in the F and S regions. 

The conductance spectrum is calculated via the
 Blonder-Tinkham-Klapwijk  (BTK) method\cite{btk} 
extended to include unconventional
superconductivity\cite{tan} and net spin polarization.
It is sufficient\cite{tan,btk} to calculate $a_S$ and $b_S$:
\begin{equation}
a_S=\frac{4 t_S L_S \Gamma_+ e^{-i\phi_{S+}}}
{U_{SS+}U_{\overline{S} S-}-
V_{SS-} V_{\overline{S} S+}
\Gamma_+\Gamma_-e^{i(\phi_{S-}-\phi_{S+})}}
\label{as}
\end{equation}
\begin{equation}
b_S=   \frac
{V_{S S+}U_{\overline{S} S-}-
U_{S S-} V_{\overline{S} S+}
\Gamma_+\Gamma_-e^{i(\phi_{S-}-\phi_{S+})}}
{U_{SS+} U_{\overline{S} S-}-
V_{SS-} V_{\overline{S} S+}
\Gamma_+\Gamma_-e^{i(\phi_{S-}-\phi_{S+})}}
\label{bs}
\end{equation}
where we introduce
$\Gamma_\pm\equiv
(\epsilon-(\epsilon^2-|\Delta_{S\pm}|^2)^{1/2})/|\Delta_{S\pm}|$, 
$L_S\equiv L_0 \cos\theta'_S/\cos \theta$, $L_0\equiv k'_F/k_F$, 
describing 
FWM, 
$t_S\equiv(1+\rho_S X)^{1/2}$, 
$t_{\overline{S}}\equiv(1-\rho_S X)^{1/2} \cos \theta_{\overline{S}}
/\cos \theta$, 
$X\equiv h_0/E_F$ , 
which defines the 
degree of polarization, 
$U_{\overline{S} S\pm}\equiv t_{\overline{S}}+w_{S\pm}$,
$V_{S S\pm}\equiv t_{S}-w_{S\pm}$,
 $w_{S\pm}\equiv L_S\pm 2iZ$, $Z\equiv Z_0/\cos\theta$, and 
$Z_0\equiv m H/\hbar k_F$  is the interfacial barrier parameter. 
The normalized differential conductance\cite{tan}
(in units of $e^2/h$) is then 
\begin{equation}
G\equiv G_{\uparrow}+G_{\downarrow}=\sum_{S=\uparrow,\downarrow} P_S 
(1+\frac{k_{\overline{S}}}{k_S}|a_S|^2-|b_S|^2)
\label{gs}
\end{equation}
where the probabilities of an incident electron with spin $S$, $P_S$, 
satisfy\cite{been},
 $P_\uparrow/P_\downarrow=(1+X)/(1-X)$. 
 At $X=0$ we recover the results of  Ref.  \onlinecite{tan}. 
The ratio of wavevectors in Eq. (\ref{gs}) reflects
that  the incident electron and the AR hole
belong to different spin bands.
One can use the conservation of 
probability current\cite{btk} to  generalize the sum rule for the
 reflection coefficients in the case of subgap conductance 
$(E<|\Delta_{S\pm}|)$, for the unpolarized case. We  get
$\frac{k_{\overline{S}}}{k_S}|a_S|^2+|b_S|^2 =1$
and then $G_S$ can be expressed in terms of 
the AR amplitude only.
For $k_{FS}>k'_F$, one sees from Eq. 
(\ref{gs}) and ``Snell's law'',
that for $|\theta|$ greater than the angle of
total reflection ($a_S=0$) $G_S=0$. We define the angularly
averaged (AA) conductance\cite{tan}, $<G_S>$, as
$<G_S>=\int_{\Omega_S} d\theta \cos \theta G(\theta) 
/\int_{\Omega_S} d\theta \cos
\theta$, where $\Omega_S$ is limited by the angle of total reflection
or by experimental setup.

We concentrate on the $d_{x^2-y^2}$ state
and take parameter values appropriate for HTSC's and ongoing
experiments\cite{amg} on the effect of spin polarization on $G$.
Typically there is a small interface resistance\cite{vas2}
i.e. 
small $Z_0$ values, away from the 
tunneling limit $Z_0 \gg 1$\cite{mers}. The experiments
use\cite{vas2} half-metallic ferromagnets with 
$X\rightarrow 1$ 
and FWM values of $L_0<1$\cite{amg}. 
We present our results for $G$ and $\langle G \rangle$ as a function of 
dimensionless energy
$E\equiv e V/\Delta_0$, where $V$ is the bias voltage. We 
take $E'_F/\Delta_0=12.5$, with $E_F=E'_F$ $(L_0=1)$ and 
$E_F=4E'_F$ $(L_0=1/2)$,
to consider the influence of FWM. 

In Fig. \ref{fig1}, we  show results for $G(E)$ at $\theta=0$,   
and $\alpha=0$ (i.e.
an  F/S interface along the (100) plane). This behavior
is the same as for an 
$s$-wave superconductor ($\Delta_{S\pm}=\Delta_0$). In panel (a), 
at zero interfacial barrier, we display the effect of
increasing $X$. 
The solid lines represent results with FWM, $L_0=1/2$. 
The behavior of the amplitude of $G$ at zero bias (AZB) reflects
the interplay between the effects of FWM and ferromagnetism.
At $X=0$  there is a ZBCD, as in 
previous work\cite{btk2}. This is caused by the effective
barrier introduced by FWM (even at $Z_0=0$), at
the interface which separates regions with different Fermi
energies. With increased exchange energy the ZBCD evolves into a ZBCP,
which narrows with decreasing $L_0$.
The ZBCP in this case is not
due to unconventional
superconductivity, since there is no 
sign change in the PP's experienced by ELQ and HLQ.
For reasonable values of the FWM, the maximum AZB is 2, 
independent of $X$. 
The AZB maximum is 
obtained from the condition
$k_\uparrow k_\downarrow=k'^2_S \equiv k'^2_F$. For
$L_0=1/2$, this occurs at $X\approx 0.968$. Thus, 
in the presence of FWM Andreev reflection can be enhanced by spin 
polarization and can even become maximal
at a special value of $X$. 
These results differ from those obtained without FWM 
when AR is always suppressed and $G$ decreases with $X$. 
The dashed lines (no FWM)
illustrate this point. In panel (b) 
we show the influence of barrier strength at fixed $X=0.6$. 
In the presence of FWM  the subgap conductance
is more reduced, with sharper peaks at $E=1$ than for the same
$Z_0$ at $L_0=1$ (dashed lines), since the effective barrier 
strength\cite{btk2} 
is enhanced. 

In Fig. \ref{fig2} we use the same parameter values and notation as in 
Fig. \ref{fig1} to display the effect of AA on $G$
for $s$-wave ($\Delta_{S\pm}=\Delta_0$ for all angles). In panel (a) the
solid curves, (with FWM) show that the ZBCP at fixed X  
 remains after AA, while its 
amplitude is typically reduced.
We see that, unlike in the unpolarized case\cite{btk2},
FWM can actually enhance $G(0)$
at fixed polarization.
In panel (b) we show that the effects of interfacial barrier on $\langle G \rangle$
are similar to those given in
Fig. \ref{fig1}.

We now turn to the effects of the sign change of the PP, i.e., of the
unconventional nature of the S region.
We take $\alpha \neq 0$ and $\theta \neq 0$ so that 
ELQ and HLQ may feel  PP's of opposite sign.
First, in Fig. \ref{fig3}, we consider the limit of
no FWM,
with $\theta =\pi/12$ and $\alpha=\pi/4$.
We examine the dependence  of the AZB
on $X$ and $Z_0$.  In panel (a) we see that
the ZBCD becomes more pronounced with higher $X$,
because  of suppressed AR. In the limit
$(E_F-h_0)\rightarrow 0$, the
subgap conductance vanishes, due to the
vanishing of the minority spin density of states (see
the bottommost curve).
For $X=0.4$ there appears a finite bias peak (FBCP), which moves to lower energy
with increasing $X$.
 This follows from 
``Snell's law''. At larger $X$, there will be increased 
difference between $k_S$ and $k'_S$, and $\theta'_S$ will depart more from 
$\theta$. ELQ and HLQ feel PP's with increasingly different spin dependent
magnitude:  $G_{\uparrow}$ and $G_{\downarrow}$ are governed 
by different energy scales. 
 When $\alpha\neq0,\pi/4$,
$G(E)$ displays two or four distinct features determined by up to four
different PP values.
In panel (b) we show the decomposition of the $X=0.6$ result
into its $G_\uparrow$ and $G_\downarrow$ components.
The position of the FBCP, discussed above, is at the
maximum of $G_\downarrow$. 
We also examine the effect of $Z_0$ at constant $X$.
With increased $Z_0$, the FBCP evolves 
towards smaller energies. Eventually, the
barrier effects 
dominate those of $X$ and the ZBCP resembles that
found in the N/S junctions, attributed to midgap surface 
states\cite{hu,tan}.

In Fig. \ref{fig4} we illustrate the effect of AA using the
parameters from Fig. \ref{fig3}. The solid curves represent AA
over all angles below total reflection and the
dashed curves are averages over a narrower region. 
Panel (a) displays the conductance for
different polarizations and $Z_0=0$.
The ZBCD reported in \onlinecite{vas2} resembles the bottommost solid curve.
The  parameter values for this curve
agree with their values in this experiment, as mentioned
above.
We see that the ZBCD in $\langle G \rangle$ occurs only at 
large $X$. The
curves  are then qualitatively different from those found in
the tunneling limit for 
an $s$-wave superconductor, where the peak
in $\langle G \rangle$ is sharp and at the gap energy.
In panel (b) we show that formation of ZBCP with increasing $Z_0$ is
a robust feature present in both types of AA.

In Fig. \ref{fig5} we consider the interplay of FWM and unconventional
superconductivity. We take $L_0=1/2$. In panel (a) we show the results
at $Z_0=0$ for a range of values of $X$. The ZBCP evolves into
a ZBCD.
In panel (b) at $X=0.7$, we see that with increasing $Z_0$ a 
FBCP forms and then evolves to a ZBCP.
The FBCP here has a different origin than the breaking 
of the time reversal symmetry state in the
 N/S system\cite{hu,tan,sauls}. From Eqs. \ref{as}--\ref{gs} 
it is simple and instructive to obtain other $G(E)$
results.

We have shown here that spin polarized tunneling spectroscopy
of F/S junctions displays qualitatively novel  behavior.
The variety of features in $G(E)$ and $\langle G(E)\rangle$ arises
from the interplay among the form of the pair potential, 
the exchange energy of 
a ferromagnet modifying the AR, and the Fermi energies and
interface properties of the F and S regions. 
The results are quite sensitive
to FWM  which  should be carefully taken into account in interpreting
spin polarized experiments and can not, in the polarized case, be simply
replaced by a change in parameter $Z_0$.

We thank A.M. Goldman, V.A. Vas'ko, K.R. Nikolaev, P.A. Kraus, 
S.W. Pierson and L. Glazman for discussions.

%
\begin{figure}
\caption{$G(E)$ (Eq. \protect{\ref{gs}}). 
$E\equiv eV/\Delta_0$. Results are for $\theta=0$ (normal incidence) and
$\alpha=0$. The solid curves are for $L_0=1/2$ (FWM present):
in panel (a) at $Z_0=0$ (no barrier) they are (from top to bottom at $E>1$)
for exchange energies
$X\equiv h_0/E_F=0,  0.6, 0.8, 0.968,0.99,0.9999$, while
in panel (b) they are at $X=0.6$ and 
(from top to bottom) $Z_0=0,0.25,0.5,1,1.5$.
The dashed curves are for
$L_0=1$ (no FWM). In panel (a) $X=0,0.6,0.968$, in
 panel (b) they are at $Z_0=0, 0.05$.}
\label{fig1}
\end{figure}
\begin{figure}
\caption{$\langle G(E)\rangle$, the $\theta$ averaged
conductivity, for the same parameter values and curve identifications 
as in Fig. \protect{\ref{fig1}}. Results in this Figure are
averaged over all angles.}
\label{fig2}
\end{figure} 
\begin{figure}
\caption{$G(E)$  
for $\theta=\pi/12$, $\alpha=\pi/4$, and $L_0=1$.
In (a), at $Z_0=0$, the curves are for 
$X=0,0.4,0.6,0.7,0.8, 0.85,0.9, 0.99$,
(top to bottom at $E=0$.) 
In (b), the solid curves correspond, from top to bottom at $E=2$, 
to $Z_0=0,0.25,0.5,1,1.5$. The dashed and dash-dotted curves
are, respectively, the $G_\uparrow$ and $G_\downarrow$
conductances at $Z_0=0$.}
\label{fig3}
\end{figure}
\begin{figure}
\caption{$\langle G(E)\rangle$ for the data in Fig. \protect{\ref{fig3}}.
The solid curves are averages over all
$\theta$ while the dashed curves are over a region of width
$\pi/24$ centered at $\pi/12$. The curves shown are, respectively,
in the same order as those in
Fig. \protect{\ref{fig3}}, except that
in (a) results for  $X=0.4, 0.7, 0.85$
have been excluded for clarity.}
\label{fig4}
\end{figure} 
\begin{figure}
\caption{$G(E)$ for $\theta=\pi/12$, $\alpha=\pi/4$, and FWM 
with $L_0=1/2$. 
In (a), $Z_0=0$,  and the curves are for 
$X= 0,0.4,0.6,0.7,0.8,0.85,0.9$ (top to bottom at $E=0$).
In (b),
the influence of $Z_0$ is shown for $X=0.7$. 
The  curves correspond, from top to bottom at $E=2$, to 
$Z_0=0,0.25,0.5,1,1.5$} 
\label{fig5}
\end{figure} 

\begin{references}
\bibitem{andre} This approximation\protect{\cite{bru}} is
valid in our region of interest at subgap conductances.
For high bias where its accuracy is reduced, $G$ is usually dominated
by extrinsic effects.
\end{references}
\end{document}